\documentclass[3p,twocolumn]{elsarticle}

\usepackage{amssymb}
\usepackage{amsmath}
\usepackage{lineno}
\journal{Chaos Solitons and Fractals}

\begin{document}


\begin{frontmatter}

\title{Tactical cooperation of defectors in a multi-stage public goods game}

\author[label1]{Attila Szolnoki}
\author[label2]{Xiaojie Chen}

\address[label1]{Institute of Technical Physics and Materials Science, Centre for Energy Research, P.O. Box 49, H-1525 Budapest, Hungary}

\address[label2]{School of Mathematical Sciences, University of Electronic Science and Technology of China, Chengdu 611731, China}

\begin{abstract}
The basic social dilemma is frequently captured by a public goods game where participants decide simultaneously whether to support a common pool or not and after the enhanced contributions are distributed uniformly among all competitors. What if the result of common efforts is {\it not} distributed immediately, but it is reinvested and added to the pool for a next round? This extension may not only result in an enhanced benefit for group members but also opens new strategies for involved players because they may act in distinct rounds differently. In this work we focus on the simplest case when two rounds are considered, but the applied multiplication factors dedicated to a certain round can be different. We show that in structured populations the winning strategy may depend sensitively on the ratio of these factors and the last round has a special importance to reach a fully cooperative state. We also observe  that it may pay for defectors to support the first round and after enjoy the extra benefit of accumulated contributions. Full cooperator strategy is only viable if the second round ensures a premium benefit of investments. 
\end{abstract}

\begin{keyword}
public goods game \sep cooperation \sep phase transition
\end{keyword}

\end{frontmatter}

\section{Introduction}
\label{intro}

According to the evolutionary selection principle, when individual interests are in conflict the most successful strategy prevails among competing approaches \cite{maynard_82}. For an adequate mathematical description we introduce a measure of success, which is a payoff value gained by a player when interacting with others. In other words, the collected payoff is a clear feedback for a player to decide on which strategy to choose. Importantly, this assumption still allows several ways how to implement the above mentioned selection principle because many alternative microscopic dynamical rules can serve this goal
\cite{amaral_pre20,yang_hx_pa19,jiao_yh_csf20}.
Beside the frequently applied imitating the more successful partner rule, we may apply alternative standards of social learning, like using myopic update, birth-death, or death-birth process, etc. \cite{szabo_jtb12,liu_rr_amc19,szolnoki_njp18b,cardoso_njp20,zhang_lm_epl19,fu_mj_pa19,szolnoki_pa18,xu_zj_c19,li_xy_epjb21,lin_jy_csf20,zhang_lm_pa21}. For further details our reader is advised to check related topical reviews \cite{szabo_pr07,perc_jrsi13,perc_pr17}.

In the standard public goods game ($PGG$), which is a simple metaphor of the conflict of two major strategies, a player decides whether to contribute to a common pool or not \cite{sigmund_10,chen_xj_srep16}. When decisions are made, we enhance the accumulated contributions by a multiplication factor and redistribute it among all group members uniformly independently of their strategies
\cite{liu_jz_epjb21,quan_j_pa21,rong_zh_c19}.
Therefore, it is not surprising that those who contribute first gain less than the defector players who just enjoy the fruit of others' efforts. This annoying conclusion cannot be avoided, just only if we assume some additional circumstance, like punishing defectors, rewarding cooperators
\cite{chen_xj_njp14,flores_jtb21,chen_xj_pcb18,lv_amc22,fu_mj_pa21,yang_hx_epl20}, or other more sophisticated mechanism which allows cooperators to collect competitive payoff \cite{cong_r_srep17,liu_jz_csf18,chen_xj_pre15,quan_j_c19,cheng_f_amc20,szolnoki_njp21,gao_sp_pre20,li_k_csf21,wei_x_epjb21,quan_j_csf21,szolnoki_amc20}. 

From our present viewpoint, however, it is more important when to realize the consequence of a chosen strategy, because players may not necessarily face with the result of their choices immediately. For example, it can happen that there is some delay when the results of strategy interactions become available
\cite{szolnoki_pre13b,yan_f_njp21}. 
Seasonal effects or long-term investments may explain why we not always evaluate strategy success immediately
\cite{szolnoki_srep19}. 
Alternatively, there are situations when the fruit of collective efforts is not distributed at once, but it is considered as an additional contribution to a new round when players are invited to cooperate to the common pool again. In this paper we study the possible consequences of such multi-stage $PGG$ by analyzing a two-level version of the classic game. This extension makes players possible to choose alternative strategies in different rounds, which enlarges the possible number of competing strategies. As an important extension, we do not insist that the enhancement factors dedicated to alternating rounds should be equal, but they could be different. This liberty helps us to reveal that the ratio of the mentioned multiplications factors has a special importance if players are arranged in a spatially structured population. Furthermore, the success of the application of mixed strategies is highly biased, because in certain parameter region it could be useful for defectors to contribute to the first round and enjoy the accumulated benefits in the last round.  

In the next section we specify the details of the extended $PGG$ and discuss the evolutionary outcome in a mixed population where all players can interact with anybody else randomly. Our key observations are summarized in Sec.~3 where we present the phase diagram of a spatially structured population and give further details of the emerging phase transitions. In the last section we briefly analyze the possible general consequences of multi-stage public goods game and briefly discuss further potential extensions. 

\section{Multi-stage public goods game}
\label{def}

By following the standard definition of $PGG$ we consider a population of $N$ players who can be cooperators ($C$) or defectors ($D$). $G$ number of players form a group to execute a collective project. While cooperators contribute a $c=1$ amount to the common pool, defectors avoid such effort. Instead, they only enjoy the benefit of the joint venture. The sum of collective investment is multiplied by an $r$ enhancement factor which expresses the fact that collective efforts result in higher income level than the simple sum of individual contributions. This enhanced amount is distributed among all group members equally, no matter if a player properly contributed to the joint effort or not. Regarding the microscopic dynamics, we apply the widely accepted pairwise comparison imitation rule. In particular, a player $y$ who has strategy $s_y$ adopts the $s_x$ strategy of a player $x$ with the following $\Gamma$ probability:
\begin{equation}
\Gamma (s_x \to s_y) = [1+\exp(\Pi_{s_y}-\Pi_{s_x})/K]^{-1}\,.
\nonumber
\end{equation}
Here $\Pi_{s_y}$ and $\Pi_{s_x}$ denote the payoff values of involved strategies, while $K$ parameter quantifies the amplitude of noise level during the adoption process. For proper comparison with previous studies of standard model we here apply $K=0.5$ value which ensures a likely adoption of more successful strategy, but also allows a reverse process with a small probability \cite{szolnoki_pre09c}.

Next we extend the standard model and introduce two-stages of the game to calculate the proper payoff values of interacting players. More precisely, when the involved group members contribute to the common pool and their investments are enhanced by an $r_1$ multiplication factor then we do not distribute this sum. Instead, we invest the whole amount into a second pool to increase the additional contributions of group members. To distinguish the potential weights of various stages, in the second round we apply $r_2$ multiplication factor to enhance the sum of players' contributions and the amount saved from the first round.
Importantly, a player may decide to contribute to the first or/and second round, hence we can distinguish four different strategies. They are designated as $DD$, $CD$, $DC$, and $CC$ depending on whether a player contributes to the first or/and to the second round, or not. 
Accordingly, the payoff value collected by different strategies from a group interaction are the following:
\begin{eqnarray}
\Pi_{DD} &=& \frac{r_2 (r_1 [n_{CD}+n_{CC}] + [n_{DC}+n_{CC}])}{G} \,,\nonumber\\
\Pi_{CD} &=& \Pi_{DD} - 1\,,\nonumber\\
\Pi_{DC} &=& \Pi_{DD} - 1\,,\nonumber\\
\Pi_{CC} &=& \Pi_{DD} - 2\,,\nonumber\\
\nonumber
\label{payoff}
\end{eqnarray}
where $n_{s}$ denotes the number of players having strategy $s$ in the group.
Here $DD$ can be considered as a ``classic" defector who never contributes to joint ventures, while a $CC$ player invests to both pools, hence can be considered as an unconditional cooperator. A key question is whether it pays to mix the attitudes by cooperating (or defect) in the first round and apply the reversed act in the second round. Or, is there a coexistence of certain strategies at specific parameter values?
 
We consider the generalized problem both in well-mixed and in structured populations. While the former is mathematically feasible in several cases, but the latter option in general is significantly closer to realistic situations
\cite{yang_gl_pa19,broom_dga21,fu_y_c21,chen_w_njp21,deng_ys_pa21,zhong_xw_pa21,zhu_pc_epjb21}
Therefore it is always instructive to compare the system behaviors for both conditions
\cite{szolnoki_epl16,roca_plr09,szolnoki_pre15}. For a proper comparison with previous results of traditional $PGG$ we here apply a square lattice topology to describe a spatially structured population where every player has four nearest neighbors, hence they form a $G=5$ member group \cite{szolnoki_pre09c}. Of course, a player is involved not only in one game, but also in four others ones which are organized by the mentioned neighbors. To gain reliable statistic we used $L \times L$ = $800 \times 800$ system size where the requested relaxation time to reach the stationary state varied between $10^4$ to $10^6$ Monte Carlo steps ($MCS$), depending on the proximity of phase transition points. According to the standard simulation protocol, during a full $MCS$ every player has a chance on average to adopt a strategy from a partner.

\begin{figure}
\centering
\includegraphics[width=7.5cm]{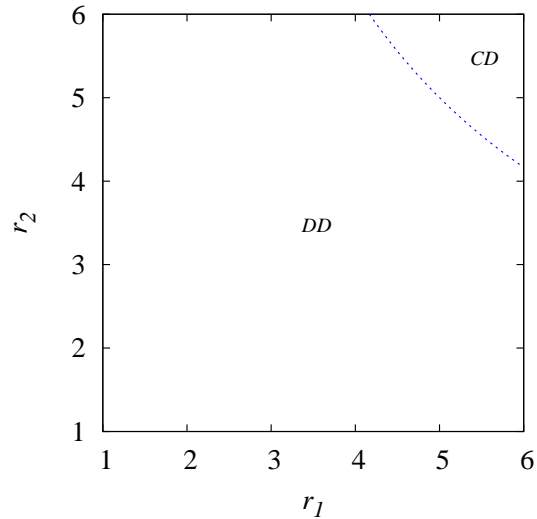}\\
\caption{Phase diagram of a mixed population on the plane of multiplication factors which are dedicated to specific rounds of $PGG$. If these factors are low, the system evolves into a trapped phase where only $DD$ players are present. If the product of $r_1 \cdot r_2$ exceeds $G \cdot G$ then $CD$ strategy prevails. These phases are separated by a discontinuous phase transition, marked by dashed blue line.}\label{mix}
\end{figure}

Before presenting the subtle system behavior of spatially structured population we first summarize the result for a well-mixed population. To make the comparison accurate we here assume that a player is also involved in $G$ games as for the spatial system. Perhaps it is worth noting that in the traditional $PGG$ defectors prevail if $r$ is lower than $G$ and cooperators can survive only above this threshold value. In our two-stage game the system behavior is conceptually similar, as shown in Fig.~\ref{mix}. Here we plotted the evolutionary stable solutions in dependence of the multiplication factors dedicated to the specific stage of the game. Similarly to the traditional one-stage game persistent defectors, i.e. $DD$ strategies are always selected if the product of $r_1 \cdot r_2$ does not exceed $G \cdot G$ threshold value. Above this point $CD$ strategy prevails.
 
In other words, when the product of enhancement factors is high enough then it pays for defector players to invest into the first round and after enjoy the accumulated benefit of the two-stage game. It is also worth noting that the liberty to choose diverse values of enhancement factor for different stages has no any relevance: the roles of enhancement factors are symmetric and the new solution can be reached independently of which multiplication factor is higher. 

\section{Phase transitions in a spatially structured population}

\begin{figure}
\centering
\includegraphics[width=7.5cm]{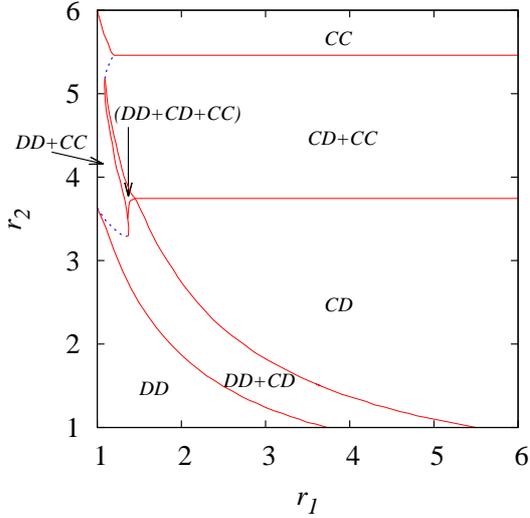}\\
\caption{Phase diagram of the system behavior in case the players are arranged on a square lattice. Solid red lines mark continuous phase transitions, while dashed blue lines denote discontinuous phase transition points. If we compare this diagram with the one obtained for well-mixed population then we can observe a significantly richer behavior where seven  evolutionary stable solutions can be found for different parameter pairs.}\label{spatial}
\end{figure}

In this section we give a detailed report about how a structured population behaves when a two-stage $PGG$ is considered. In evolutionary game theory we frequently observe that the system behavior is strikingly different when we leave the analytically feasible well-mixed condition and assume an interaction topology where players have fixed and limited number of partners 
\cite{lutz_pre21,li_qr_pa20,zheng_lp_pa21}.
This is the case in our present model, too, as it is demonstrated in Fig.~\ref{spatial}. 

The first observation is the large diversity of stable solutions on the parameter plane. Namely, we can detect seven different phases as we change the ($r_1,r_2$) parameter pair. Similarly to the well-mixed case $DD$ strategy becomes dominant for small values of enhancement factors, but this phase is significantly smaller here. The tactical cooperation of defectors, in other words the $CD$ strategy, can also be the winner of the evolutionary process, but only if $r_2$ value remains moderate. Furthermore, persistent cooperation of $CC$ strategy could be dominant for high $r_2$ values independently of the value of $r_1$. This feature underlines the strong asymmetry in the role of enhancement factors, which cannot be detected in a well-mixed population. 

Beside the mentioned pure phases we can observe coexistence of different strategies. For example $DD$ and $CD$ strategies can form a stable solution between the related pure phases. Similar behavior can be observed for the relation of $CD$ and $CC$ strategies or the coexistence of the extreme strategies of $DD$ and $CC$. Interestingly, $DD$, $CD$, and $CC$ strategies can form a stable three-member solution at specific pairs of enhancement factors. But $DC$ strategy is not proved to be viable for any combination of parameters.  

\begin{figure}
\centering
\includegraphics[width=8.0cm]{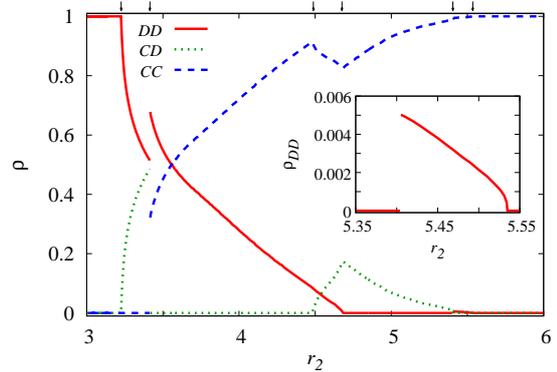}\\
\caption{The portion of surviving strategies in dependence of $r_2$ enhancement factor as we cross the phase diagram at a fixed $r_1 = 1.16$ value. The gradual change of $r_2$ results in six(!) consecutive phase transitions whose positions are marked by arrows on the top. The last two transition points are hardly visible on this scale therefore we zoomed the portion of $DD$ strategy in the inset. The error bars are comparable to the thickness of the lines.}\label{vert}
\end{figure}

The mentioned phases are separated by continuous (marked by solid red) or discontinuous phase transition points. The latter borders are denoted by dashed blue lines. To illustrate the richness of system behavior in Fig.~\ref{vert} we present a cross-section of the diagram obtained at $r_1=1.16$ as we gradually increased the value of $r_2$. By solely varying $r_2$ we can detect six consecutive phase transitions. Their positions are marked by arrows on the top. The first transition happens at $r_2 = 3.226$ when $CD$ becomes viable and coexist with $DD$ strategy due to the relatively high value of $r_2$. After, at $r_2 = 3.415$ an interesting first-order phase transition happens, where a qualitatively new solution, the coexistence of $DD$ and $CC$ emerges. Later we will give further details which explains this discontinuous phase transition.
 
As we increase $r_2$ further, $CD$ strategy becomes viable again and forms a stable three-member solution with the mentioned strategies from $r_2 = 4.488$. However, as the phase diagram highlights, this solution is restricted to a very narrow area of parameters because it requires a not too large $r_1$ value, otherwise $CD$ would enjoy its first investment too easily. In parallel, $r_2$ should be at an intermediate value: for smaller $r_2$ values $DD$ would dominate, while a high second enhancement factor would help cooperator strategies. This is why at $r_2 = 4.676$ cooperator strategies, $CD$ and $CC$ become more efficient and form a more successful two-strategy solution again. Lastly, we can observe a rather exotic reentrant to $DD+CC$ phase, but this recovery of the mentioned solution is strongly related to the low value of $r_1$. For larger $r_1$ values, when the first round is more effective, this solution cannot be detected anymore, but gives way to the $CD$ tactical defector strategy. Staying at the mentioned exotic phase transitions, the first one is discontinuous at $r_2 = 5.404$ while the second one is continuous at $r_2 = 5.535$, as it is illustrated in the inset of Fig.~\ref{vert} where we zoomed the portion of $DD$ strategy for clarity.
 
\begin{figure}
\centering
\includegraphics[width=8.0cm]{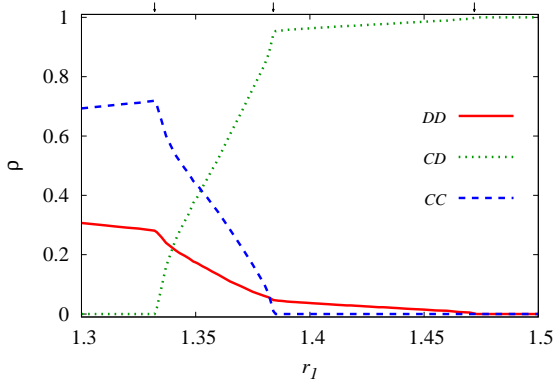}\\
\caption{The portion of surviving strategies in dependence of $r_1$, as we cross the phase diagram at fixed $r_2 = 3.7$ value. Three phase transitions can be detected as we reach the full $CD$ phase at high $r_1$ values. Their positions are marked by arrows. The error bars are comparable to the thickness of the lines.}\label{hor}
\end{figure}

Of course, we can detect a series of phase transitions as we increase the benefit of first round at a fixed $r_2$ value, but the system behavior is simpler than previously. A representative cross-section is given in Fig.~\ref{hor} where we plot the portion of surviving strategies as a function of $r_1$ at a fixed $r_2 = 3.7$ value. In the low $r_1$ region only the extreme strategies, $DD$ and $CC$, can survive. This fact can be interpreted that at very low $r_1$ the system becomes practically identical to a classical, single-stage $PGG$, where only pure strategies can be viable. Their relations depends only on the value of $r_2$, which takes the role of the traditional enhancement factor, $r$. When it is low then only $DD$ survives, while high $r_2$ can provide the full domination of $CC$. Between these extreme cases the mentioned strategies coexist due to the spatiality of the population. This is the case at $r_2 = 3.7$, as we mentioned above. One may note that this value is below the $r = 3.745$ threshold value of classic model \cite{szolnoki_pre09c}, but of course here the first round may contribute a bit which lowers this critical point modestly. Turning back to Fig.~\ref{hor}, as we increase $r_1$, $CD$ becomes viable and form the previously mentioned three-member solution with their partners. By supporting the first round further, $CD$ becomes really effective and crowds out classic $CC$ strategy. In other words, it pays defectors to invest into the first round because later they can harvest a significantly higher payoff at the end. Figure~\ref{hor} also shows that this tactical defector strategy becomes exclusive at a relatively small $r_1$ value by giving no space even for $DD$ strategy. The phase diagram shown in Fig.~\ref{spatial} also highlights that this strategy remains viable for the majority of ($r_1, r_2$) parameter pairs and only very low $r_1$ or very high $r_2$ is capable to replace it with a traditional strategy.

\begin{figure}
\centering
\includegraphics[width=5.5cm]{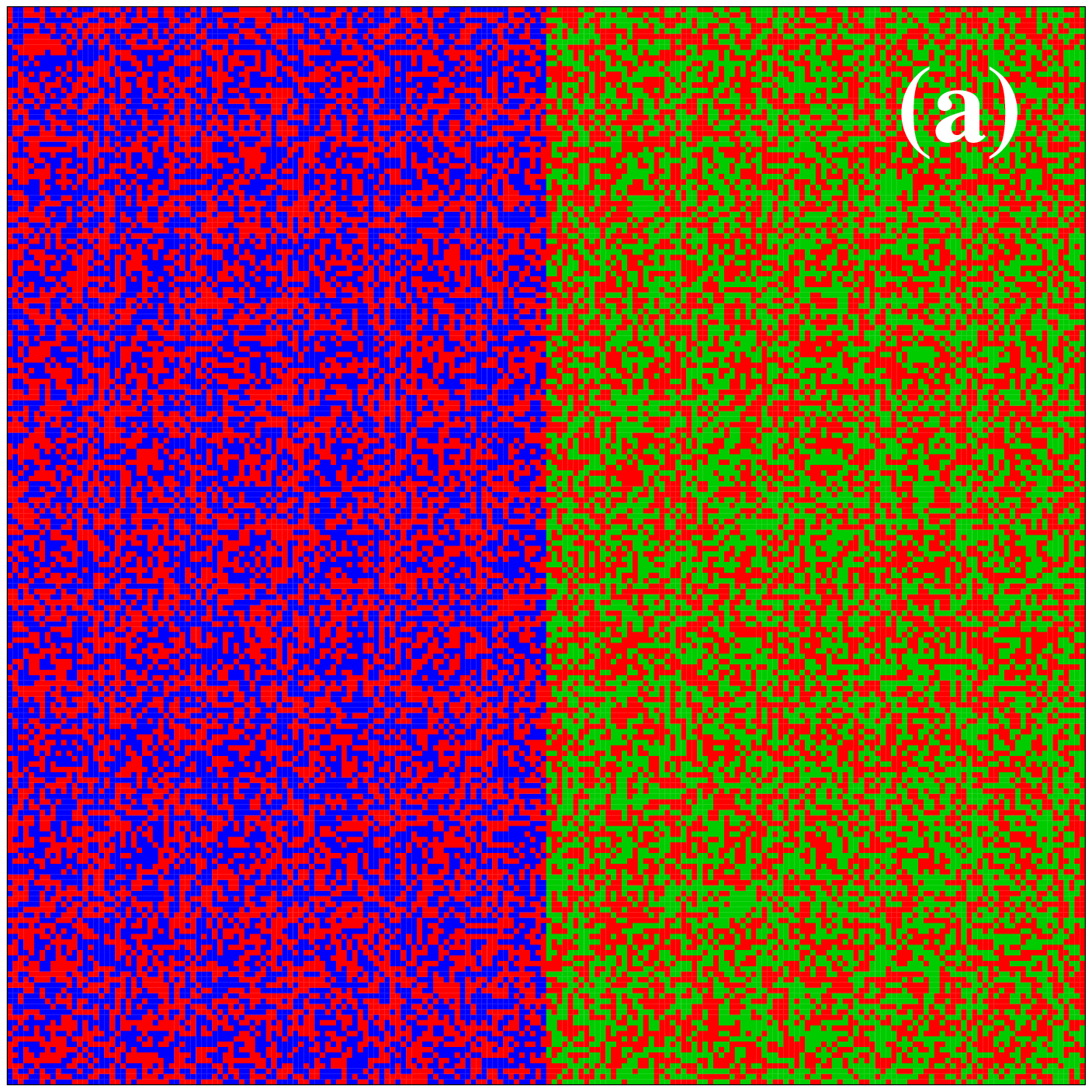}\\
\includegraphics[width=5.5cm]{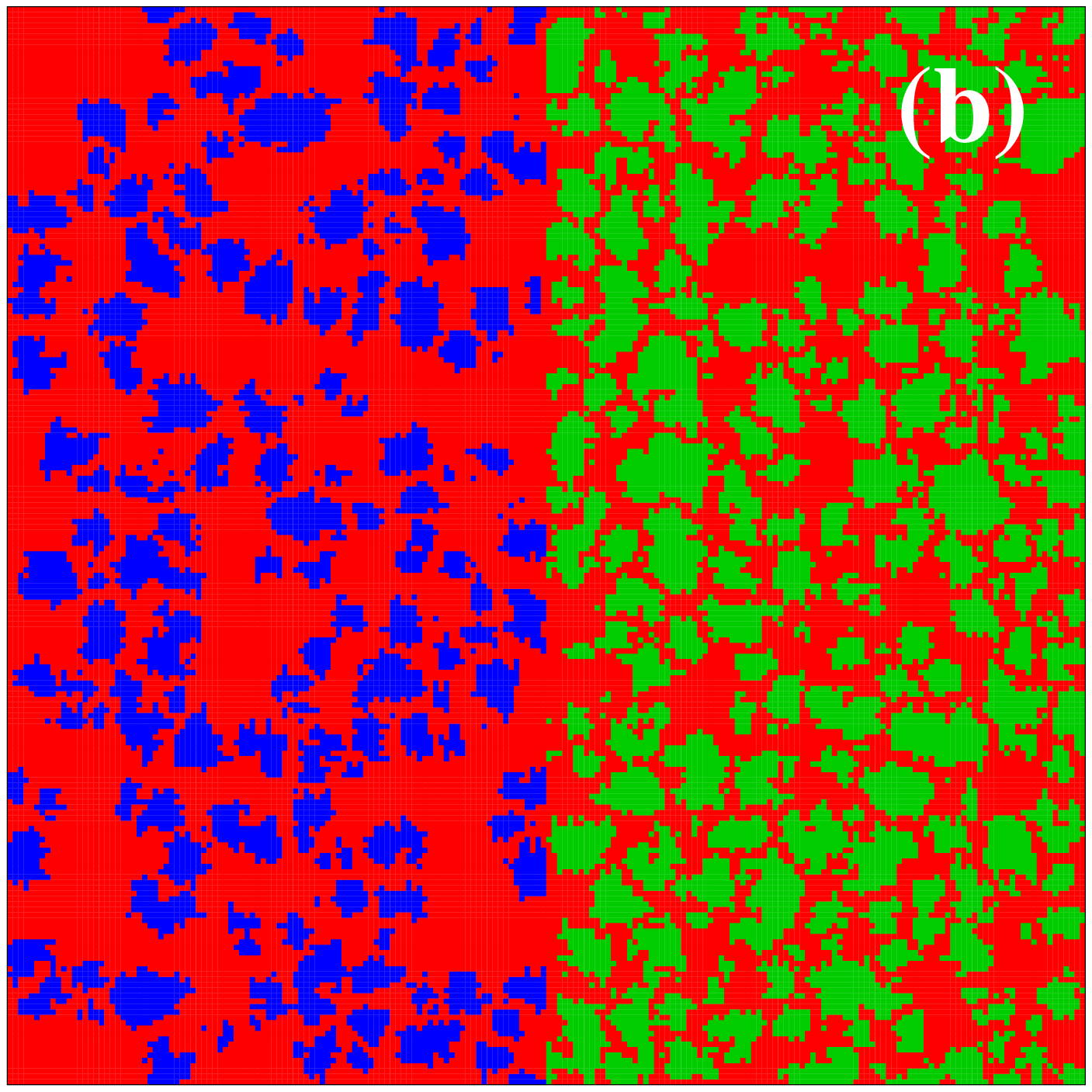}\\
\includegraphics[width=6.0cm]{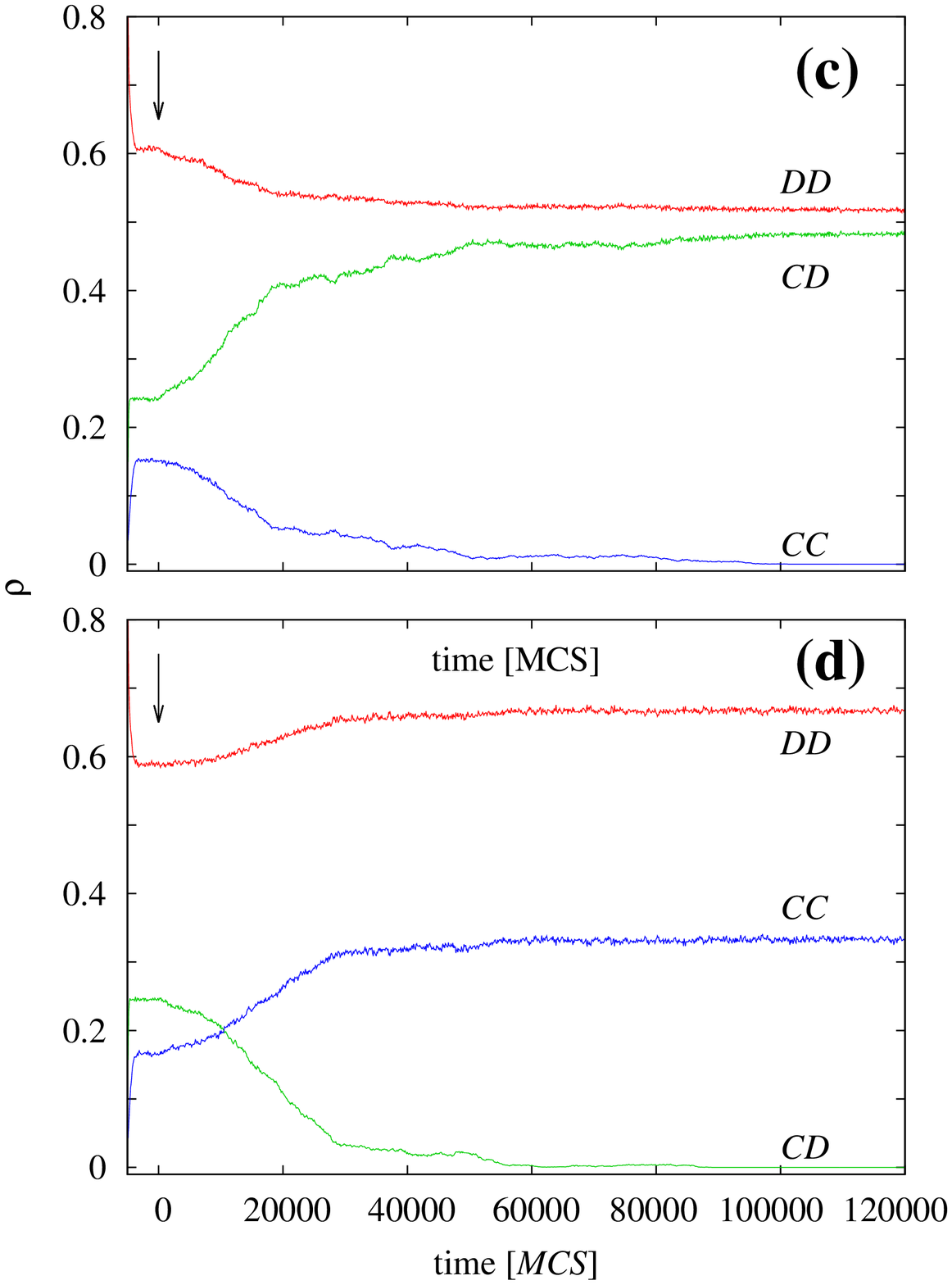}\\
\caption{Stability analysis of sub-solutions to explain the origin of discontinuous phase transitions obtained at $r_2 = 3.41$ (c) and $r_2 = 3.42$ (d) in a $800 \times 800$ system. In both cases $r_1 = 1.16$. Further details and more specific explanation can be found in the main text.}\label{first}
\end{figure}

In the following we give a deeper insight about the origin of discontinuous phase transitions we previously reported. For this reason we monitor the strategy evolution at two parameter pairs which are very close to each other, but they are in different sides of the transition point. More precisely, we follow the evolution at $r_1 = 1.16$, $r_2 = 3.41$ and at $r_1 = 1.16$, $r_2 = 3.42$. To make our point more visible, we use a prepared initial state where we divide the available space into two halves where the first half contains $DD$ and $CC$ players randomly, while the second half is occupied by $DD$ and $CD$ players. This initial setup is illustrated in panel~(a) of Fig.~\ref{first}. 

Here, in agreement with the color code used in Fig.~\ref{vert} and in Fig.~\ref{hor}, we marked by red color $DD$ players, while blue (green) color denotes $CC$ ($CD$) players on the grid. After we launch the evolution according to the two-stage $PGG$, but imitations are allowed only within the subsystems we defined. As a result, in both halves of the space a stationary solution emerges. The mentioned stage is illustrated in Fig.~\ref{first}(b). This panel highlights that both $DD+CC$ and $DD+CD$ could be a destination of the evolutionary process. We note that the presented patterns were obtained at $r_2 = 3.41$, but practically we would get identical strategy distribution for $r_2 = 3.42$ by following similar protocol.
These solutions can be reached after 5000 $MCS$ of relaxation. In the following we open the borders for imitations and allow the solutions to compete for space. Accordingly, either $DD+CC$ or $DD+CD$ will dominate and invade the whole space depending on the actual value of $r_2$. These final destinations are not shown here but they show similar patterns as the sub-system solutions depicted in panel~(b). 

The trajectories of the reported evolutions are shown in panel~(c) and panel~(d), where $r_2 = 3.41$ and $r_2 = 3.42$ values were used respectively. Here the random initial states of the subsystems were started at -5000 $MCS$, and the borders were opened at zero time, which are marked by arrows in these panels. As panel~(c) illustrates, the portion of $CC$ strategy starts falling immediately, while it grows when we gently modified the $r_2$ value in panel~(d). Finally the system evolves into the mentioned two-member solutions when the third strategy dies out. We note that we used $800 \times 800$ system size to avoid too noisy trajectories, but panel~(a) and (b) show only just a $200 \times 200$ portions around the borders, otherwise the compact domains of $CC$ and $CD$ strategies would be less visible. 

This comparison underlines that both solutions would be a stable outcome in the absence of the other one. Technically, both $CC$ and $CD$ players form compact islands in the sea of $DD$ players, as it can be seen in panel~(b). This is how network reciprocity mechanism works in general. Namely, these convex patterns make the cooperating strategies possible to reach competitive payoff despite of the relatively low values of enhancement factors. When we are at the phase transition point exactly then these solutions are in equilibrium. But if we leave this point slightly then this equilibrium is broken and one of the solutions dominates the other. Notably, in the vicinity of this transition point the relaxation is slow, as it is illustrated in panel~(c) and (d) in Fig.~\ref{first}, because the driving force to reach the final state is weak. But this destination is inevitable if we wait enough in a necessary large system. Summing up, what we can observe here are the typical features which characterize first-order phase transitions in statistical physics.

\begin{figure}
\centering
\includegraphics[width=7.5cm]{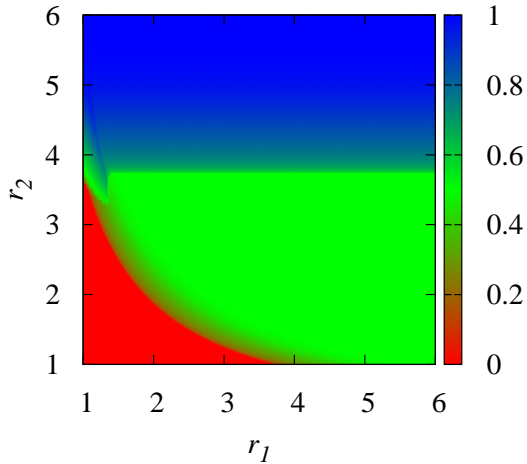}\\
\caption{Heat map of cooperation level on the parameter plane. Red color denotes the fully defector state while blue color marks full cooperation. Green color shows an intermediate state when players partly cooperate on average during the two-stage process. Legend on the right quantifies these levels. The asymmetry how the final outcome depends on the values of enhancement factors is striking.}\label{coop}
\end{figure}

Last we turn back to the comparison of outcomes obtained in well-mixed and in structured populations. As we already noted, a significant difference can be detected in the roles of enhancement factors. While in a random system it has no particular significance which factor is larger because their product counts, in structured population this symmetry is seriously broken. To support our argument in Fig.~\ref{coop} we show the average cooperation level of players during the two-stage game. In agreement with previously used color coding we here mark by red color those parameter pairs where players never cooperate in the stationary state. Similarly, blue color denotes those parameter values when players behave as classical cooperators and contribute to both pools. Green color shows those $(r_1, r_2)$ pairs when there is one contribution on average. Perhaps it is worth noting that this average can be reached in different ways: a mixture of $DD$ and $CC$ players or pure $CD$ players can provide similar levels of cooperation. The main point here is the strong asymmetry of heat map in terms of the values of enhancement factors. In particular, the value of $r_1$ could be really high, but if $r_2$ is modest, we can only reach an intermediate cooperation level. On the other hand, if $r_2$ is high enough then we can reach the desired high cooperation level almost independently of the value of $r_1$. 

\section{Conclusions}

It is always an individual decision whether we want to cooperate and contribute to a common venture or not. In short term it could pay to save such cost and only enjoy the fruit of efforts, but in long term, if everybody else think similarly then we cannot avoid the so-called tragedy of the commons trajectory
\cite{hardin_g_s68,cheng_f_pa19,szolnoki_epl11,liu_dn_pa19,tao_yw_epl21,szolnoki_pre10b,yang_lh_csf21}. 
In other words, we will face to the consequence of our choice. But this outcome should not be clear immediately if we are part of a multi-stage project and the success of our choice, namely the collected payoff will only be clear at the very end. A delayed evaluation or the consequence of a multi-level project, however, opens a new direction for players to apply more sophisticated strategies than the classic unconditional cooperator and defector choices. For example, they may cooperate or defect in the early stage and later they may alter their attitudes. In this work we explored the consequence of a two-stage public goods game, where the fruit of the first round is not distributed among the players, but it is reinvested into the second stage. It is an important extension that the weight of different rounds could be unequal, which also opens to study the combination of heterogeneous joint ventures.

We have shown that the proposed extension has no particularly interesting consequence in a well-mixed population because it simply rescale the enhancement factors. Therefore only unconditional defectors survive if the product of enhancement factors remains below a threshold value. The only interesting feature is emergence of tactical defectors who dominate above the mentioned threshold value. They only invest onto the first round, and enjoy the accumulated benefit.  The roles of enhancement factors dedicated to a certain round remain symmetric and only their product counts.

This simple picture becomes more complex in structured populations. We here presented the results obtained on square grid which makes also possible to compare the results with traditional model played on similar topology. The presented phase diagram revealed several phases which are formed by a pure strategy or the combination of these. We could detect not only two-member coexistence, but also a solution in which three strategies can survive in the stationary state. These phases are separated by continuous or discontinuous phase transition lines. We have given explanation why the latter can emerge in this system.

Perhaps the most spectacular feature of the spatially structured population is the strong symmetry breaking regarding the roles of enhancement factors dedicated to certain rounds. A moderate level of cooperation can be reached at relatively low $r_2$ value if the $r_1$ value is high enough, but complete cooperation can only be reached for a high $r_2$ value. In the latter case, however, the value of $r_1$ has no particular importance.
This observation may help us how to design multi-stage games because the last round has paramount importance to reach a high cooperation level.

To follow this path we may study what if more stages are present in a $PGG$ where the results of previous rounds are reinvested again and again until the last stage when all cards are revealed. Of course, such an extension makes the study really challenging because the number of the available strategies grows exponentially with the number of rounds. Furthermore, if we allow that distinct stages may influence the final outcome in different ways, i.e. we introduce stage-specific enhancement factors, then the optimal player's strategy could be more interesting. For example, successful players may want to defect, cooperate and defect again. Hopefully, following works will clarify this research avenue. 

\vspace{0.5cm}

This research was supported the National Natural Science Foundation of China (Grants No. 61976048 and No. 62036002).

\bibliographystyle{elsarticle-num-names}

\end{document}